\documentclass[11pt]{article}
\usepackage[dvips]{graphicx}

\begin{document}
\title{\Large \bf Neutrino capture by r-process waiting-point nuclei}
\author{ \\ Rebecca Surman and Jonathan Engel  \\
{\small Dept.\ of Physics and Astronomy, CB3255,
University of North Carolina, Chapel Hill, NC 27599  }}

\maketitle
\begin{abstract}
{\normalsize We use the Quasiparticle Random Phase Approximation to include 
the effects of low-lying Gamow-Teller and first forbidden strength in neutrino 
capture by very neutron-rich nuclei with $N$ = 50, 82, or 126.  For electron 
neutrinos in what is currently considered the most likely $r$-process site the 
capture cross sections are two or more times previous estimates.  We briefly
discuss the reliability of our calculations and their implications for 
nucleosynthesis.}
\end{abstract}

\newpage

The {\it r} process, which is responsible for the formation of half of
all elements with $A>70$, is thought to take place in the ``hot bubble" that
expands off a proto-neutron star during a type II supernova\cite{Woo}.  If
that is the case the nuclei involved are subject to an intense
neutrino flux while they are made.  This fact has been used to
try to constrain the distance of the $r$-process site from the neutron 
star\cite{Ful} and to argue that the time scale for the process is less than 
usually believed\cite{Qian1}.  

Serious investigation of these and related issues will require knowledge of
cross sections for neutrino capture by the very neutron-rich nuclei that lie
along the $r$-process path.  Although several groups have recently estimated
the cross sections\cite{Ful,McL,Qian}, none has tried to be very precise, for
two reasons.  First, precision is difficult unless one focuses on a few
nuclei, and the $r$-process involves a huge number of neutron-rich isotopes.
Second, conditions in the hot bubble are so uncertain that precise estimates
are not really warranted.  Here, nonetheless, we attempt to calculate
charge-changing neutrino-nucleus scattering rates\footnote{The corresponding
antineutrino rates are suppressed in neutron-rich nuclei and won't be
considered here.}  a little more carefully than before in particularly
important ``waiting-point" isotopes at neutron closed shells.  The reason is
not so much to be precise as it is to investigate effects that systematically
increase cross sections above current estimates.

Uaing the neutrino-nucleus scattering formalism presented in Ref.\ \cite{Wal},
we examine two kinds of nuclear transitions not yet considered in this context
that can be prompted by neutrinos and therefore add to the cross section:
Gamow-Teller (GT) and first-forbidden transitions to {\it low-lying} excited
states.  Since most of the GT transition strength, induced by the operator
$\vec{\sigma} \tau_+$, lies in a single broad resonance the low-energy
strength has usually been neglected; Ref.\ \cite{Qian}, for example,
approximates the entire distribution by a Gaussian of width 5 MeV.  But
the energy of the GT resonance in very neutron-rich nuclei is probably high
enough to prevent excitation by most hot-bubble neutrinos, the average energy
of which is only about 11 MeV.  The same is true of the isobar analog (IA)
resonance, which is excited even less because $J=0$ states have only a single
$M$-substate.  The low-lying GT strength may therefore contribute nearly as
much to cross sections as the resonances, even though it is small in
comparison.

Forbidden transitions induced by the operators $\vec{r} \tau_+$ and
$[\vec{r}\vec{\sigma}]^{J=0,1,2} \tau_+$ have not been considered either,
except briefly and without definite conclusion in Ref.\ \cite{McL}, because in
stable nuclei they too are concentrated in high-lying (dipole-like) resonances
and, moreover, are further suppressed by a factor of $(qR)^2$, where $R$ is
the nuclear radius.  In very neutron-rich nuclei, however, it is possible that
some forbidden strength lies low.  The reason is that the forbidden operators,
which change parity and ordinarily must a create a proton one oscillator shell
(roughly speaking) above the neutron they destroy, can actually create a
proton in an oscillator shell below the destroyed neutron if there are many
more neutrons than protons.  The diagram in figure 1 shows that in the
unstable nuclei with $N=82$, for example, operators with $J^\pi=1^-$ and $2^-$
can transform a neutron near its Fermi surface to a proton close to its Fermi
surface.  As a result, some forbidden strength may lie low enough to overcome
the $(qR)^2$ suppression (which is really not so great in heavy nuclei) and
contribute significantly to the neutrino-nucleus cross sections.  The likely
magnitude of this unusual contribution, and of that due to low-lying GT
strength, is what we examine here.

\begin{figure}[htb]
\begin{center}
\includegraphics[width=10cm,height=11cm]{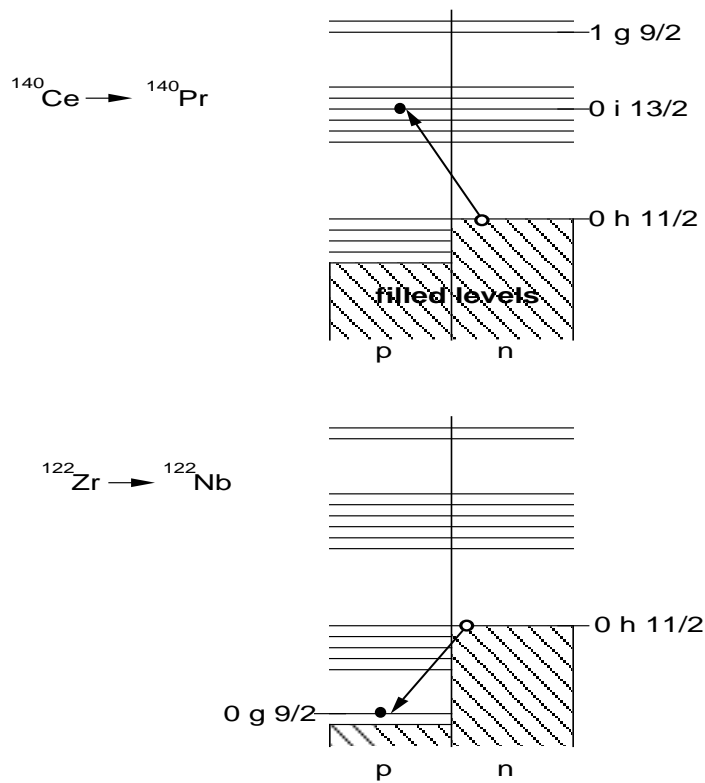}
\end{center}
\caption{Single-particle picture of forbidden charge-changing (neutron
$\longrightarrow$ proton) transitions in a stable nucleus and a 
neutron-rich nucleus, both with $N=82$.  The change in parity restricts 
neutrons to move up an oscillator shell as they become protons in 
the stable nucleus, but in the neutron-rich nucleus they can move down an 
oscillator shell as well.}
\end{figure}

Assessing the effects of states outside a giant resonance requires a
calculation sophisticated enough to represent the competition between
single-particle structure, which is responsible for the existence of low-lying
strength, and the residual nucleon-nucleon interaction, which gathers strength
into the higher-lying resonance.  At the same time, and for reasons mentioned
above, it is probably not worthwhile to aim for high precision.  We therefore
use a relatively simple microscopic description based on the neutron-proton
Quasiparticle Random Phase Approximation (QRPA)\cite{Hal,Cha}.  We restrict
ourselves to nuclei with closed neutron shells ($N=50,82,126$) in part to
avoid dealing with deformation but also because the closed-shell nuclei form
``bottlenecks" in the {\it r}-process flow and thus determine the time scale
of nucleosynthesis.  The nuclear-structure effects that are significant here
probably play a role in deformed nuclei as well.

Our calculations proceed as follows:  We obtain single-particle energies from
the parameterized Wood-Saxon potential in Ref.\ \cite{Ber}.  Where possible we
include all neutron and proton levels that participate with reasonable
probability in transitions induced by the Gamow-Teller and forbidden
operators.  Those that are unbound we take to be resonances; to obtain their
wave functions we neglect the Coulomb interaction, making them bound.  (We
include Coulomb effects in perturbation theory when calculating their
energies.)  This procedure results in single-particle spaces of up to 20
levels for each kind of particle.  Within these large spaces we use two
different residual two-body interactions in the QRPA.  The first is a
$\delta$-function with independent strengths in the particle-hole and
particle-particle channels as described in Ref.\ \cite{Eng}, and the second a
combination of 7 Yukawa potentials fit in Ref.\ \cite{Hos} to a
Paris-potential G-matrix.

To reproduce systematics in stable nuclei and extrapolate to neutron-rich
isotopes we modify both forces.  First we adjust the strength of each
interaction in the pairing channel to obtain pairing gaps $\Delta_p = \Delta_n
= 12/\sqrt{A}$ MeV\cite{Hom} from the BCS equations.  In the RPA we then
adjust 2 parameters, the strengths of the particle-hole interactions in the
$0^+$ and $1^+$ channels, so as to place the IA and GT resonances at
appropriate energies.  Unfortunately, although the energy $E_{IA}$ of the
analog state follows from the Coulomb energy difference between parent and
daughter nuclei, the appropriate value of $E_{GT}$ in very neutron-rich nuclei
is less certain.  (It is also more important, since the GT resonance
contributes more to the cross section.)  The usual way to estimate $E_{GT}$ is
through a relation of the form $E_{GT}=E_{IA}+\delta$, where, $\delta$ has a
linear dependence on neutron excess.  Though one can fit this parameter to
data in the valley of stability\cite{Nak}, it is not obvious how best to
extend it to unstable nuclei.  Fuller and Meyer\cite{Ful} take $\delta$ to be
0 for $Z/A<0.377$, while Qian, et al.\ \cite{Qian} extrapolate the fit from
Ref.\ \cite{Nak}, $\delta = 26A^{-1/3}-18.5 (N-Z) A^{-1}$ MeV, with the result
that $E_{GT}<E_{IA}$ far from stability.  Here we try both prescriptions.  We
also use $g_A=1$ for the GT transitions to account for missing strength.

Not surprisingly, as is apparent from the top half of figure 2, the
$\delta$-function and G-matrix based interactions produce similar GT
distributions\footnote{The large resonance widths are due to a prescription
for spreading taken from Ref.\ \cite{Smith}; they may not be realistic but
altering them has little effect on the neutrino cross sections.}  once they
have been modified/fit as just described, and in particular both predict
significant amounts of low-lying strength (though noticeably less than in
stable nuclei for $N$ = 82 and 126).  The reason we use two interactions is
really the uncertainty in the location of the forbidden strength, which is
harder to measure, decompose, and fit than allowed strength.  Instead of
relying on the sketchy systematics that do exist\cite{Hor}, we use the two
forces without further modification to get a handle on forbidden transitions.
For the $\delta$-function interaction, the strengths in the $IA$ and $GT$
channels determine the strengths of the $S=0$ and $S=1$ parts of the force,
and we fix these once they have been fit to the allowed resonances.  For the
G-matrix based interaction, we use the original negative-parity proton-neutron
matrix elements.  As shown in the bottom of the figure, the $2^-$ strength
distributions produced by the two interactions are quite similar.  Both forces
in the end do predict some low-lying forbidden strength; the large neutron
excess prevents the giant resonances from swallowing everything up.  As a
result the forbidden transitions as well as the GT transition to low-lying
states can change the reaction rates substantially.

\begin{figure}[htb]
\begin{center}
\includegraphics[width=12cm,height=10cm]{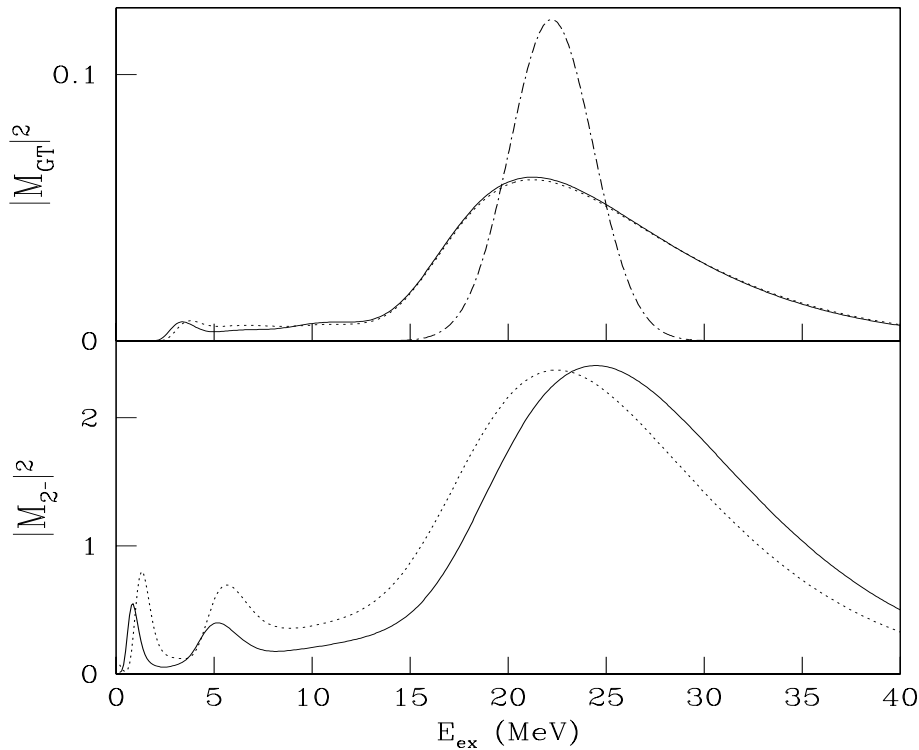}
\end{center}
\caption{Calculated GT (a) and forbidden $2^-$ (b) strength distributions for 
$^{124}$Mo.  The solid lines come from the $\delta$-function interaction
and the dashed lines from the G-matrix based interaction.  The dot-dashed line 
in (a) is from a Gaussian GT resonance centered at the energy of the isobar 
analog state and with a width of 5 MeV, normalized to give the correct total 
strength.}
\end{figure}

To see this quantitatively, we use the (approximate) spectrum of supernova 
electron neutrinos\cite{Qian},
\begin{equation}
f(E_\nu) = {1\over F_2(\eta_\nu)T_\nu^3}{E_\nu^2 \over 
\exp[(E_\nu/T_\nu)-\eta_\nu]+1}~,
\label{eq:1}
\end{equation}
where $T$ is the temperature, $\eta_{\nu}$ the chemical potential, and $F_2$ 
a normalizing factor, to obtain spectrum-averaged neutrino cross sections
\begin{equation} 
\langle\sigma_\nu\rangle=\int f(E_\nu) \sigma(E_\nu) dE_\nu ~.
\end{equation}
The capture rates are directly proportional to $\langle\sigma_\nu\rangle$.
Several values for the chemical potential $\eta_\nu$ appear in the literature,
but while changes in that parameter affect the overall rates, they don't
change the relative importance of low-lying strength with respect to the
resonant strength considered previously.  We therefore simply use $\eta_\nu=3$
and adjust $T$ so that the average neutrino energy is about 11 MeV.

Contributions of the low-lying GT and forbidden strength to the
spectrum-averaged cross sections for two representative nuclei appear in
figures 3 and 4.  These particular plots result from the $\delta$-function
interaction and the prescription $E_{GT}=E_{IA}$, but the notable features are
always the same.  Figure 3 shows the contribution of low-lying GT strength in
$^{78}$Ni and $^{190}$Gd to $f(E_\nu) \sigma(E_\nu)$ as a function of neutrino
energy.  The solid line represents the cross sections obtained with the full
QRPA GT strength distribution calculated here, while the dashed line
represents those calculated as in Ref.\ \cite{Qian}, with a Gaussian GT
strength distribution centered at the energy of the isobar analog state and
having a width of 5 MeV (the GT distributions themselves are compared in
figure 2).  The low-lying strength increases the cross sections for neutrinos
below 15 MeV substantially, especially when the neutron excess is very large.
We may even be underestimating the extent of the increase.  The GT
distribution shown in figure 1 has considerably less low-lying strength than
do most stable nuclei.  The reason is that near the drip line the resonance is
higher in energy and particle-hole force must therefore be stronger, with the
side effect that more GT strength is pulled into the resonance.  If the GT
distributions at $N$ = 82 and 126 looked more like those in stable nuclei the
allowed contribution to neutrino rates would go up even further.

\begin{figure}[htb]
\begin{center}
\includegraphics[width=12cm,height=10cm]{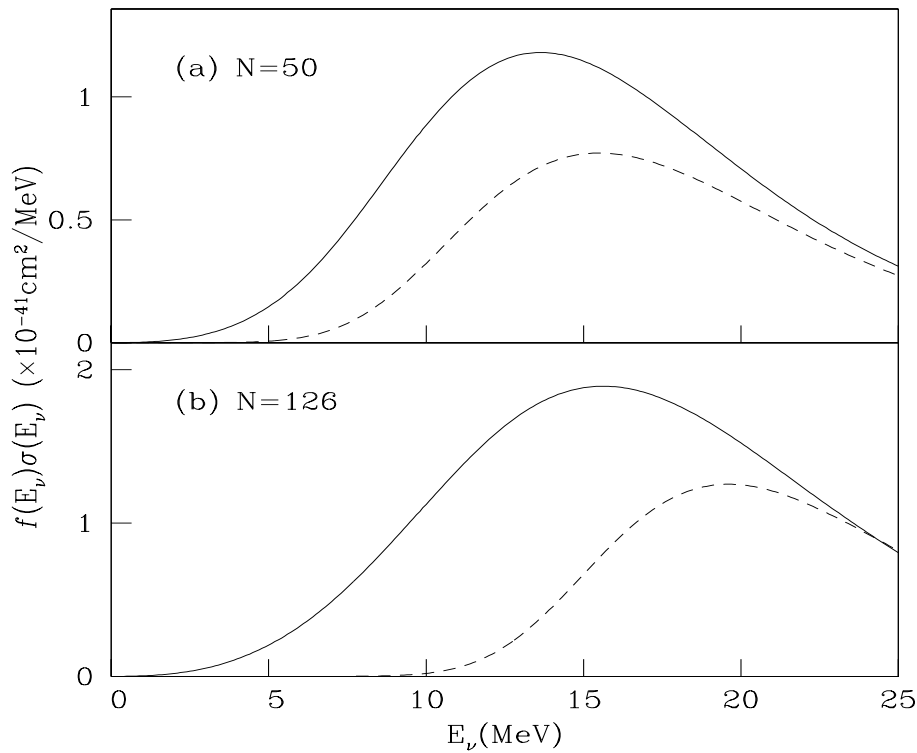}
\end{center}
\caption{GT contributions to spectrum-averaged neutrino-nucleus scattering 
cross sections for (a) $^{78}$Ni and (b) $^{190}$Gd.  The solid line is
the full GT contribution calculated here and the dashed line  the
contribution of the Gaussian GT distribution.}
\end{figure}

Figure 4 shows the role played by forbidden transitions in the same two
nuclei; it compares the total GT contribution to $f(E_\nu)\sigma(E_\nu)$ just
discussed with the contributions of the $1^-$ and $2^-$ transitions.  The
low-lying forbidden strength ought to increase with neutron excess and,
indeed, figure 4 shows that the forbidden portion of the cross section in
$^{190}$Gd is clearly larger than that in $^{78}$Ni.  In general, the unusual
low-lying forbidden strength provides about 5--10\% of the total
spectrum-averaged cross section when $N=50$, 10--20\% when $N=82$, and
20--35\% when $N=126$.

\begin{figure}[htb]
\begin{center}
\includegraphics[width=12cm,height=10cm]{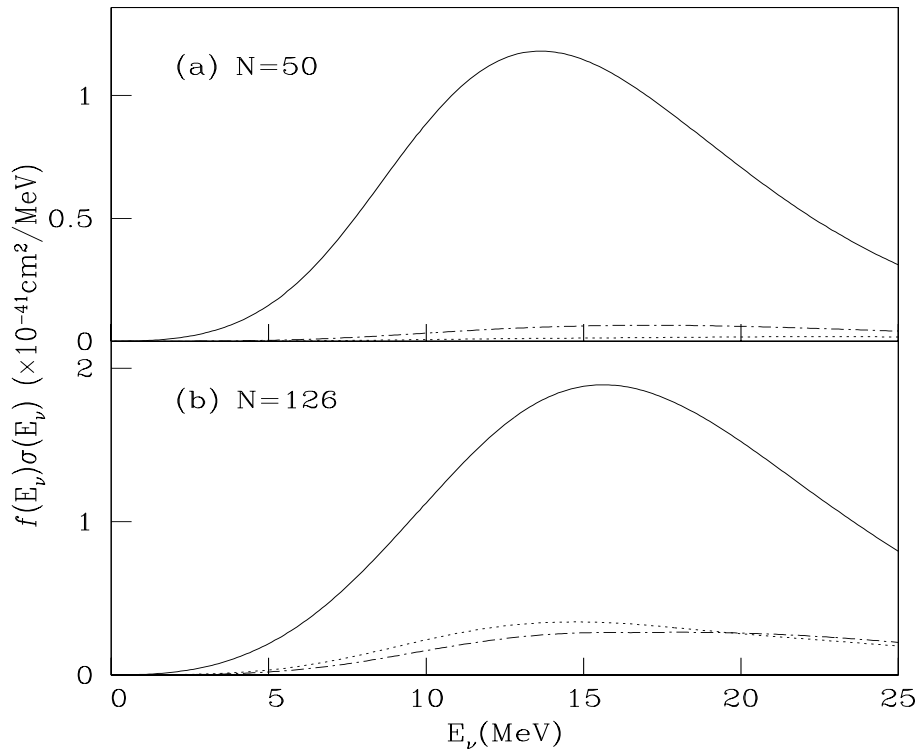}
\end{center}
\caption{Contributions to  spectrum-averaged neutrino-nucleus scattering cross 
sections for (a) $^{78}$Ni and (b) $^{190}$Gd.  The solid line is the GT
contribution and the dashed and dot-dashed lines are the $1^-$ and $2^-$ 
first-forbidden contributions.}
\end{figure}

Table 1 compares $\langle\sigma_\nu\rangle_{\rm total}$, our total cross
section, with $\langle\sigma_\nu\rangle_{0}$, that calculated under the
assumption that only the IA state and GT resonance (with width 5 MeV)
contribute, in several nuclei.  The factors in the table are calculated with
$E_{GT}=E_{IA}$; rates with $E_{GT} < E_{IA}$ are all of course larger, but
the ratio of $\langle\sigma_\nu\rangle_{\rm total}$ to
$\langle\sigma_\nu\rangle_{0}$ remains roughly constant for the $N=50$ and
$N=82$ nuclei.  (For the $N=126$ nuclei, it drops by 10--20\% because the
parameter $\delta$, which lowers the GT resonance, grows rapidly with neutron
excess for large $A$).More detailed results for three representative nuclei
and the several prescriptions described above appear in table 2, where
$\langle\sigma_\nu\rangle_{\rm total}$ is broken up into allowed and
forbidden parts.  In the heaviest nuclei our estimates are more than twice as
large as those made before, and we believe it more likely that they are too
small than too large.

\newpage
\begin{center}
\begin{table}[htb]
\begin{center}
\caption{Comparison of total spectrum-averaged $\nu_e$ cross sections as 
calculated in this work, $\langle\sigma_\nu\rangle_{\rm total}$, to 
allowed-only cross sections, $\langle\sigma_\nu\rangle_0$, calculated with
the Gaussian GT strength distribution (see text). All cross sections are in 
units of $10^{-41} {\rm cm}^2$.}
\begin{tabular}{cccc}
Z & N & A & $\langle\sigma_\nu\rangle_{\rm total}/\langle\sigma_\nu\rangle_0$ 
\\ \hline
26 & 50 & 76 &  1.6 \\
28 & 50 & 78 &  1.6 \\
30 & 50 & 80 &  1.6 \\ \hline
40 & 82 & 122 &  1.8 \\
42 & 82 & 124 &  1.8 \\
46 & 82 & 128 &  1.8 \\
48 & 82 & 130 &  1.7 \\ \hline 
62 & 126 & 188 &  2.3 \\
64 & 126 & 190 &  2.2 \\
66 & 126 & 192 &  2.2 \\
68 & 126 & 194 &  2.1 
\end{tabular}
\end{center}
\end{table}

\begin{table}[hbt]
\caption{Total spectrum-averaged $\nu_e$ cross sections for three 
representative nuclei, with two different forces and two prescriptions for the 
position of the GT centroid.  All cross sections are in units of
$10^{-41} {\rm cm}^2$.}
\begin{tabular}{ccccccccc}
Z & N & A & prescription & force & $\langle\sigma_\nu\rangle_{0}$ & 
$\langle\sigma_\nu\rangle_{\rm allowed}$ & 
$\langle\sigma_\nu\rangle_{\rm forbidden}$ & $\langle\sigma_\nu\rangle_{\rm 
total}$ \\ \hline 
28 & 50 & 78 & $E_{GT}=E_{IA}$ & $\delta$-function & 13.6 & 19.9 & 1.5 &
21.4 \\
   &  &  &  & G-matrix & 13.6 & 22.4 & 1.4 & 23.8 \\
   &  &  & $E_{GT}=E_{IA}+\delta$ & $\delta$-function & 16.1 & 24.1 & 1.8
& 25.9 \\
   &  &  &  & G-matrix & 16.1 & 26.6 & 1.4 & 28.0 \\
42 & 82 & 124 & $E_{GT}=E_{IA}$ & $\delta$-function & 19.6 & 29.4 & 4.9 &
34.3 \\
   &  &  &  & G-matrix & 19.6 & 30.1 & 7.0 & 37.1 \\
   &  &  & $E_{GT}=E_{IA}+\delta$ & $\delta$-function & 29.7 & 41.3 & 6.3
& 47.6 \\
   &  &  &  & G-matrix & 29.7 & 42.0 & 7.0 & 49.0 \\
64 & 126 & 190 & $E_{GT}=E_{IA}$ & $\delta$-function & 20.6 & 34.3 & 11.9 
& 46.2 \\
   &  &  &  & G-matrix & 20.6 & 38.9 & 21.3 & 60.2 \\
   &  &  & $E_{GT}=E_{IA}+\delta$ & $\delta$-function & 35.0 & 49.0 &
14.7 & 63.7 \\
   &  &  &  & G-matrix & 35.0 & 49.4 & 21.3 & 70.7    
\end{tabular}
\end{table}
\end{center}


To sum up, we find that low-lying GT and first forbidden strength increases
the rate of neutrino scattering from very neutron-rich nuclei, usually by
factors of two or more.  Our cross sections are still uncertain, in part
because of our relatively poor understanding of nuclei far from stability.
More careful and accurate calculations are possible even without more data,
but in our view the effort is not warranted yet because of the still
quite large uncertainty in, e.g., the flux of hot-bubble neutrinos.  No matter
what the sources and amounts of uncertainty, however, our results point to a
{\it uniform} and significant increase in capture rates over previous
estimates.  Any serious calculation must include the effects we describe, and
neutrino capture will play a larger role in the $r$-process than one would
otherwise believe.



\begin{thebibliography}{99}

\bibitem {Woo} S.E.\ Woosley, J.R.\ Wilson, G.J.\ Mathews, R.D.\ Hoffman, 
and B.S.\ Meyer, Astrophys.\ J.\ {\bf 433}, 229 (1994).

\bibitem{Ful} G.M.\ Fuller and B.S.\ Meyer, Astrophys.\ J.\ {\bf 453}, 792 
(1995).

\bibitem{Qian1} Y.-Z.\ Qian, Nucl. Phys. A {\bf 621}, 363 (1997).

\bibitem{McL} G.C.\ McLaughlin and G.M.\ Fuller, Astrophys.\ J.\ 
{\bf 455}, 202 (1996).

\bibitem{Qian} Y.-Z.\ Qian, W.C.\ Haxton, K.\ Langanke, and P.\ Vogel,
Phys. Rev. C {\bf 55}, 1532 (1997).

\bibitem{Wal} J.D.\ Walecka, in: Muon Physics, Vol. 2, eds. V.W.\ Hughes 
and C.S.\ Wu (Academic, New York, 1975), p. 113.

\bibitem{Hal} J.A.\ Halbleib and R.A.\ Sorensen, Nucl. Phys. A {\bf 98}, 542 
(1967).

\bibitem{Cha} D.\ Cha, Phys. Rev. C {\bf 27}, 2269 (1983).

\bibitem{Ber} G.F.\ Bertsch, The Practitioner's Shell Model (American 
Elsevier, New York, 1972).

\bibitem{Eng} J.\ Engel, P.\ Vogel, and M.R.\ Zirnbauer, Phys. Rev. C {\bf 
37}, 731 (1998).

\bibitem{Hos} A.\ Hosaka, K.-I.\ Kubo, and H.\ Toki, Nucl. Phys. A {\bf 444}, 
76 (1985).

\bibitem{Hom} H.\ Homma, E.\ Bender, M.\ Hirsch, K.\ Muto, H.V.\
Klapdor-Kleingrothaus, and T.\ Oda, Phys. Rev. C {\bf 54}, 2972 (1996). 

\bibitem{Nak} K.\ Nakayama, A.\ Pio Galeao, and F.\ Krmpotic, Phys. Lett. 
B {\bf 114}, 217 (1982).

\bibitem{Smith} R.D. Smith and J. Wambach, Phys. Rev. C {\bf 38}, 100 (1988). 

\bibitem{Hor} D.J.\ Horen, C.D.\ Goodman, D.E.\ Bainum, C.C.\ Foster,
C.\ Gaarde, C.A.\ Goulding, M.B.\ Greenfield, J.\ Rapaport, T.N.\
Taddeucci, E.\ Sugarbaker, T.\ Masterson, S.M.\ Austin, A.\ Galonsky, W.\
Sterrenburg, Phys. Lett. B {\bf 99}, 383 (1981).

\end{thebibliography}
\end{document}